\documentclass[aps,
twocolumn,
showpacs,prl]{revtex4}
\usepackage{graphicx}
\usepackage{latexsym}
\usepackage{amsfonts}
\usepackage{amssymb}
\usepackage{amsmath}
\begin{document}
\title{Conformational phase diagram for polymers adsorbed at ultrathin nanowires}
\author{Thomas Vogel}
\email[E-mail: ]{t.vogel@fz-juelich.de}
\author{Michael Bachmann}
\email[E-mail: ]{m.bachmann@fz-juelich.de}
\homepage{http://www.smsyslab.org}
\affiliation{Soft Matter Systems Research Group, Institut f\"ur Festk\"orperforschung (IFF-2), 
Forschungszentrum J\"ulich, D-52425 J\"ulich, Germany}
\begin{abstract}
We study the conformational behavior of a polymer adsorbed at
an attractive nanostring and construct the complete structural phase diagram
in dependence of the binding strength and effective thickness of the string.
For this purpose, Monte Carlo optimization techniques are employed to
identify lowest-energy structures for a coarse-grained hybrid
polymer--wire model. Among the representative conformations in the 
different phases are, for example,
compact droplets attached to the string and also nanotube-like monolayer films
wrapping the string in a very ordered way. We here systematically analyze
low-energy shapes and 
structural order parameters to elucidate the transitions between the structural phases.
\end{abstract}
\pacs{82.35.Gh,05.10.Ln,61.48.De}
\maketitle
The interaction of polymers with substrates is relevant in nanoscale
applications like molecular nanoelectronic circuits and in biological
processes such as receptor--ligand binding. Thus, scrutinizing basic
structural mechanisms of molecular binding at interfaces is crucial
for a large field of interdisciplinary research and for potential
applications.

In recent years, there has been substantial progress in understanding
general properties of polymer adhesion at solid substrates. This
includes, for example, the identification of generic structural phases
and the transitions between
these~\cite{vrbova1,milchev01jcp,singh1,bj4,prellberg1,bj6,linse1,paul1,mbj1},
as well as specific binding affinities of proteins regarding the type
of the substrate and the amino acid
sequence~\cite{bj6,whaley1,sarikaya1,goede1}.  In most of these
studies, the substrate is considered as being planar. The influence of
curved substrates on the formation of structural phases has been
subject of works on droplet and helix formation at
cylinders~\cite{milchev02jcp,srebnik07cpl}.

In this Letter, we systematically study conformational phases induced
by an attractive nanowire, i.e., a substrate with one-dimensional
topology.  A nanowire could be, for example, a stretched polymer with
the ends attached to dielectrical beads fixed by optical tweezers.  It
is one of the most striking results of our study that under certain
parametrizations of the polymer--nanowire interaction, i.e., in the
corresponding region of the conformational phase diagram, the polymer
crystallizes in stable cylindrical shapes with monomer alignments
which resemble atomic arrangements known from single-walled carbon
nanotubes.  In this conformational phase, the cylindrical hull
surrounds the thin wire such that the interior is free of particles. In
a hydrodynamic application, for example, molecules can still flow
through it. Since the axis of a polymeric tube is always oriented
parallel to the direction of the wire, the growth direction of the
tube can be controlled. This would enable the construction of complex
tube systems and, therefore, allows for applications beyond those
known for conventional nanotubes. Another conceivable application of
this structural coincidence is the \emph{systematic} stabilization or
functionalization of nanotubes by polymer
coating~\cite{hasan09am,valentini02japs,joan08prb}.

In our study, we investigate a coarse-grained model for the polymer
and a linelike substrate representing the nanowire. For the polymer,
we employ a linear bead-stick model, i.e., covalent bonds between the
$N$ monomers are stiff. The chain is not grafted to this string and may move
freely.
The total energy of the system includes three contributions,
$E=E_\mathrm{LJ}+E_\mathrm{bend}+E_\mathrm{string}$.  The interaction
between nonadjacent monomers is governed by the standard Lennard-Jones
(LJ) potential,
\begin{equation}
E_\mathrm{LJ}(\{r_{ij}\})=4\epsilon_\mathrm{m}\sum_{i=1}^{N-2}\sum_{j=i+2}^{N}\left[\left(\frac{\sigma_\mathrm{m}}{r_{ij}}\right)^{12}-\left(\frac{\sigma_\mathrm{m}}{r_{ij}}\right)^{6}\right],\label{eq:1_lj}
\end{equation}
with the distance $r_{ij}$ between nonbonded monomers $i$ and $j$. The 
monomer--monomer interaction parameters $\epsilon_\mathrm{m}$ and $\sigma_\mathrm{m}$ are set to
unity in the following. The weak bending energy is a remnant of the protein-like origin of the 
model~\cite{stilli93pre} and reads
$E_\mathrm{bend}(\{\cos\theta_i\})=\kappa\sum_{i=2}^{N-1}(1-\cos\theta_i)$
with the bending stiffness set to $\kappa=1/4$. The bending angle $\theta_i$ is defined by the 
covalent bonds connected to the $i$th monomer. The monomer--string energy is obtained by continuously integrating 
a standard LJ potential over the infinitely long string~\cite{vogel10xxx}. We find
\begin{equation}
E_\mathrm{string}(\{r_{\perp;i}\})=\pi\, a\epsilon_\mathrm{f}\sum_{i=1}^N\left(\frac{63}{64}\frac{\sigma_\mathrm{f}^{12}}{r_{\perp;i}^{11}}-\frac{3}{2}\frac{\sigma_\mathrm{f}^{6}}{r_{\perp;i}^5}\right),
\end{equation}
where $\sigma_\mathrm{f}$ and $\epsilon_\mathrm{f}$ are the monomer--string
interaction parameters. The distance of the $i$th monomer perpendicular to the string is denoted by
$r_{\perp;i}$. For convenience, we scale the potential such that its minimum
is
$-1$ at $r_{\perp}^\mathrm{min}$ for
$\epsilon_\mathrm{f}=1$ and $\sigma_\mathrm{f}=1$, in which case
$a\approx 0.528$~\cite{vogel10xxx}. 
The effective thickness of the string, $\sigma_\mathrm{f}$, is related
to the minimum distance $r_{\perp}^\mathrm{min}$ of the monomer--string
potential via
$r_{\perp}^\mathrm{min}(\sigma_\mathrm{f})=\left({693}/{480}\right)^{1/6}\sigma_\mathrm{f}\approx1.06\,\sigma_\mathrm{f}$.
Alternatively, the monomer--string energy can be considered as the limiting case
of the interaction of a monomer with a cylinder (cp. Ref.~\cite{milchev02jcp}) of
radius $R\to 0$, keeping the overall LJ ``charge'' fixed~\cite{vogel10xxx}.

\begin{figure}
\centerline{
\includegraphics[width=\columnwidth]{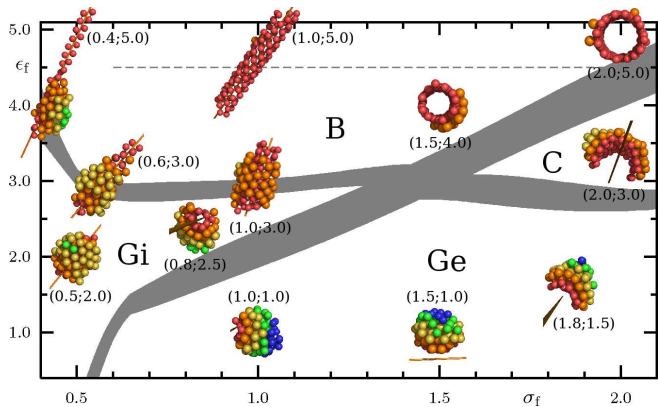}}
\caption{The conformational phase diagram parametrized by the monomer--string
potential parameters; from left to right, the effective string thickness $\sigma_\mathrm{f}$
increases and from bottom to top, the string attraction strength
$\epsilon_\mathrm{f}$ gets larger. Gray bands (widths correspond to
uncertainty) indicate transition lines between compact, crystalline polymer
structures with the string inclosed (Gi) or excluded (Ge),
crescent-shaped (C), and barrel-like (B) conformations. The dashed
line indicates a topological crossover that separates mono- and
multilayer regions. Inset pictures show representative
low-energy states. Monomers with the same coloring (or shadings) belong to the same 
layer.}
\label{fig:1}
\end{figure}

We now systematically analyze the
conformational phases of a polymer with $N=100$ monomers
for different values of effective thickness and attraction strength of the string,
$\sigma_\mathrm{f}$ and $\epsilon_\mathrm{f}$, 
respectively. By ``phase'' we denote a domain in the parameter space,
where the representative conformations share
qualitatively the same morphology. We have convinced ourselves in exemplified simulations that
the results we discuss in the following for the 100mer are also qualitatively correct 
for longer chains. In Fig.~\ref{fig:1},
the conformational phase-diagram is shown and representative
adsorbed polymer structures are depicted. This phase diagram is a result
of extensive analyses of structural properties for more than~$150$
low-energy conformations with different parametrizations.
For the identification of lowest-energy conformations, stochastic generalized-ensemble Monte Carlo
methods~\cite{bergneuh91plb} and deterministic
conjugate-gradient optimization were used. Conformational changes were
performed by applying a
variety of update moves, including local crankshaft, slithering-snake,
global spherical-cap, and translation moves~\cite{vogel10xxx}. 

In the phase diagram, four major structural phases can be identified.
For weak attraction ($\epsilon_\mathrm{f}\lesssim3$), two types of
crystalline droplets~\cite{svbj1} adhered to the string (regions Ge,
Gi) can clearly be distinguished. Either the string axis is inclosed
inside the droplet (Gi) or passes by externally (Ge).  If the adhesion
strength of the string $\epsilon_\mathrm{f}$ increases, compact
droplets in Gi melt near $\epsilon_\mathrm{f}\approx 3$ and phase B is
entered, where polymer conformations extend along the string axis.
Near $\epsilon_\mathrm{f}\approx 4.5$, a~crossover (dashed line in
Fig.~\ref{fig:1}) from the multilayer barrel structures to monolayer
conformations with strong similarities to single-walled nanotubes
occur. According to former studies of polymer adsorption at planar
substrates~\cite{bj4,mbj1}, this crossover corresponds to a
topological transition between three-dimensional compact crystalline
and two-dimensional filmlike structures. For sufficiently large values
of the effective string thickness $\sigma_\mathrm{f}$, the polymer
layers do not completely wrap the tube and stable crescent-shaped
``clamshell-like''~\cite{milchev01jcp} structures dominate in the
region denoted by C.
\begin{figure}[t]
 \includegraphics[width=\columnwidth]{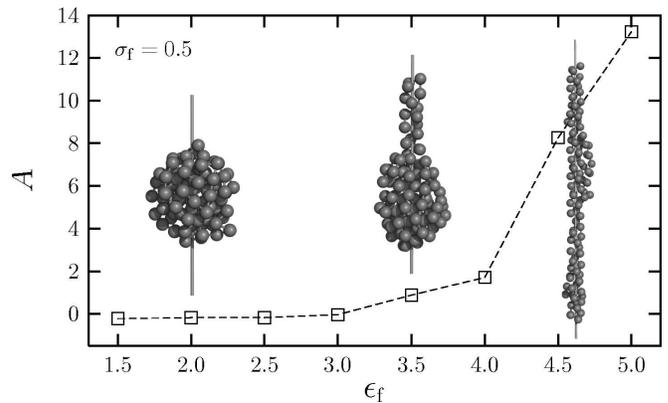}
 \caption{Asymmetry parameter $A$ at small effective string thickness
 ($\sigma_\mathrm{f}=0.5$). The inset pictures show corresponding
 conformations at $\epsilon_\mathrm{f}=2.0$, $3.5$, and $4.5$, illustrating
 the transition from spherical-symmetric to stretched
 cylindrical-symmetric low-energy structures. The conformational
 transition between Gi and B occurs near
 $\epsilon_\mathrm{f}=3.0$. For larger values of
 $\epsilon_\mathrm{f}$, $A$ starts to significantly deviate from zero
 and conformations become cylindrical.}
 \label{fig:2}
\end{figure}

Let us now have a closer look at the different conformational transitions. 
In region Gi,
compact crystalline conformations with spherical symmetry dominate. 
Increasing in this regime the attraction strength
$\epsilon_\mathrm{f}$ while keeping the effective thickness $\sigma_\mathrm{f}$ fixed, 
this symmetry breaks at
$\epsilon_\mathrm{f}\approx 3$ and the cylindrical phase B is entered. 
This transition can be best
characterized by 
introducing an asymmetry parameter based on the gyration tensor components
parallel and perpendicular to the string,
$A=r_\parallel^\mathrm{gyr}/r_\perp^\mathrm{gyr}-1$. This order parameter is shown as a function of
$\epsilon_\mathrm{f}$ in Fig.~\ref{fig:2}, exemplified 
for $\sigma_\mathrm{f}=0.5$. In the spherical regime Gi, 
$A\approx 0$. As expected,
$A$ increases for $\epsilon_\mathrm{f}\gtrsim 3$ and the structures become asymmetric. 
At this point,
it is equally favorable for a monomer to stick to the string, or to
form contacts to neighboring monomer layers. The conformations
stretch along the string until they form a maximally compact
monolayer tube surrounding the string for
$\epsilon_\mathrm{f}\gtrsim 4.5$.
\begin{figure}[t]
\includegraphics[width=\columnwidth]{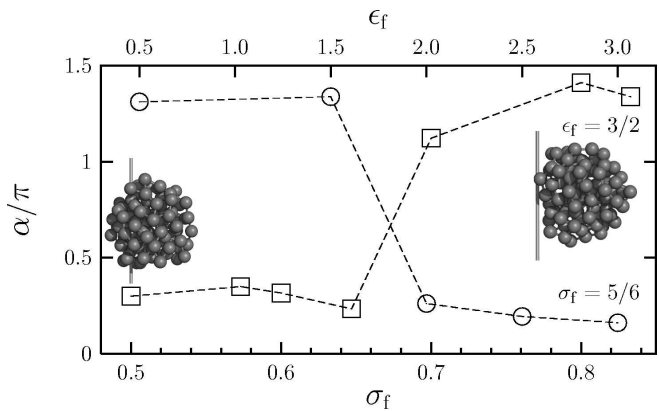}
\caption{The plot shows the opening angle $\alpha$ of low-energy
conformations for $\epsilon_\mathrm{f}=3/2$ in dependence of $\sigma_\mathrm{f}$ 
(squares, lower scale)
and for $\sigma_\mathrm{f}=5/6$ as a function of $\epsilon_\mathrm{f}$
(circles, upper scale). The inset pictures show corresponding
conformations at $\epsilon_\mathrm{f}=3/2$ and $\sigma_\mathrm{f}=1/2,4/5$,
illustrating the separation of the droplet from the string.}
\label{fig:3}
\end{figure}

Increasing, on the other hand, the effective thickness
$\sigma_\mathrm{f}$ for values of the attraction strength
$\epsilon_\mathrm{f}< 3$, the transition from Gi to Ge is
characterized by the different locations of the string relative to the
droplet: it is inclosed (Gi) or excluded (Ge). For small values of
$\epsilon_\mathrm{f}$, the transition point can be estimated as a
first approximation by assuming a tetrahedral monomer-packing in the
crystalline droplet. Then, the circumsphere radius of a tetrahedron is
$r_\circ\approx 0.61$ which corresponds to a limiting effective string
thickness $\sigma_{\mathrm{f},\circ}\approx 0.58$. Thus, inserting a
string with $\sigma_\mathrm{f} < \sigma_{\mathrm{f},\circ}$ does not
break intra-monomer contacts within a compact structure. Above this
limiting value, however, the string would cause an energetically
disfavored replacement of monomers inside the conformation and is
hence ``pushed'' out of the droplet.

Quantitatively, this
transition can be identified by measuring
the opening angle $\alpha$ of a given conformation. Projecting the positions of monomers in contact 
with the string onto a plane perpendicular to the
string, $\alpha$ is defined as
the angle between the string and two monomers that spans 
the largest region of the plane with no monomers residing
in. Thus, roughly, conformations with $\alpha< \pi$ correspond to conformations inclosing the
string (Gi), whereas $\alpha >\pi$, if the string is located outside the droplet (Ge).
Figure~\ref{fig:3} shows how $\alpha$ changes when crossing the 
transition line Gi$\leftrightarrow$Ge horizontally or vertically.
Fixing the attraction strength at $\epsilon_\mathrm{f}=3/2$, $\alpha$ increases 
rapidly from $\sigma_\mathrm{f}\approx 0.65$ (squares, lower scale), which is close to the 
estimate $\sigma_{\mathrm{f},\circ}$ given above.
For larger 
values of the effective thickness
the string is shifted outwards to retain optimal monomer packing.
The inset pictures show lowest-energy conformations at $\epsilon_\mathrm{f}=3/2$
and $\sigma_\mathrm{f}=1/2$ (representative 
for phase Gi) and $\sigma_\mathrm{f}=4/5$ (Ge), respectively.
Increasing, on the other hand, the string attraction strength $\epsilon_\mathrm{f}$
while $\sigma_\mathrm{f}=5/6$ is fixed (circles, upper scale), the inclusion
of the string, accompanied by a rapid decrease of $\alpha$, occurs at $\epsilon_\mathrm{f}\approx 1.75$.

\begin{figure}[t]
\includegraphics[width=\columnwidth]{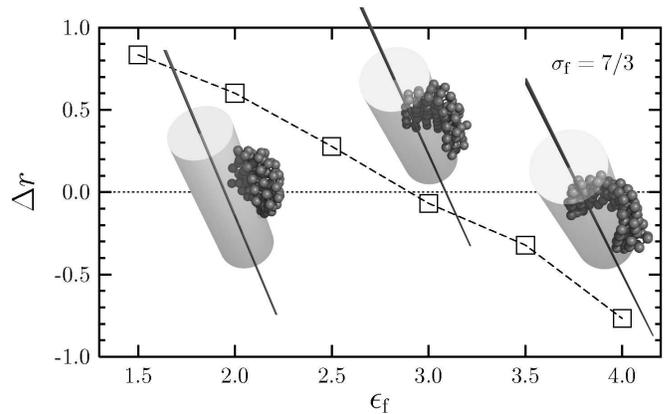}
\caption{Distance $\Delta r$ of the center of mass of the polymer 
from the virtual surface of the cylinder with the radius that corresponds to the
minimum position of the string potential ($r_\perp^\mathrm{min}\approx 1.06\sigma_\mathrm{f}$) for
$\sigma_\mathrm{f}=7/3$. The intersection of the curve with the dotted line ($\Delta r=0$)
at $\epsilon_\mathrm{f}=2.9$, where the center of mass equals the radius of this cylinder, defines the
transition from Ge to C. Pictures show conformations at $\epsilon_\mathrm{f}=2$, $3$, and~$4$.}
\label{fig:4}
\end{figure}

Starting in Ge and increasing $\sigma_\mathrm{f}$ and $\epsilon_\mathrm{f}$ above certain
threshold values results in the transition towards adsorbed curved
conformations (C) in the sense that the polymer begins to wrap 
the string. Different monomer layers form. 
We quantitatively define this transition to occur at the
point, where the distance of the center of mass of the polymer from the
string, $r_\perp^\mathrm{com}=N^{-1}|\sum_{i=1}^N \vec{r}_{\perp,i}|$, 
equals the monomer--string potential minimum distance 
$r_\perp^\mathrm{min}(\approx 1.06\sigma_\mathrm{f})$, i.e., at $\Delta
r=r_\perp^\mathrm{com}-r_\perp^\mathrm{min}=0$. 
Qualitatively, the center of mass intrudes into the virtual cylinder with radius $r_\perp^\mathrm{min}$,
defined by the inner layer of monomers. In
Fig.~\ref{fig:4}, $\Delta r$ is plotted as a function of $\epsilon_\mathrm{f}$ at
$\sigma_\mathrm{f}=7/3$. The transition point $\Delta r=0$ is marked
by the dotted line which is here intersected at $\epsilon_\mathrm{f}= 2.9$, in correspondence
to the Ge$\leftrightarrow$C transition line in the phase diagram in Fig.~\ref{fig:1}. 
The shown inset pictures represent conformations 
with $\Delta r=0.6,-0.1,-0.8$ at $\epsilon_\mathrm{f}=2,3,4$.\looseness1

\begin{figure}
\includegraphics[width=\columnwidth]{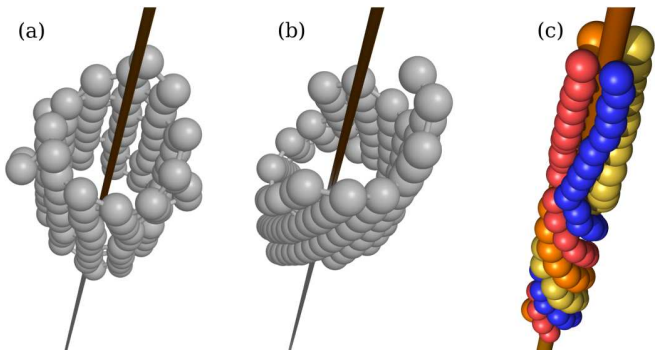}
\caption{Highly ordered cylindrical monolayer
conformations of adsorbed polymers with different wrappings in the barrel phase B 
at $\epsilon_\mathrm{f}=5$ 
for (a) $\sigma_\mathrm{f}=1.50$, (b)
$\sigma_\mathrm{f}=1.57$, and (c) $\sigma_\mathrm{f}=0.65$ (different colors or shadings 
shall facilitate the perception only). Geometric properties of these structures resemble 
chiral alignments of atomic structures known from single-walled nanotubes.}
\label{fig:5}
\end{figure}

Finally, increasing $\epsilon_\mathrm{f}$ further, region B is
entered, i.e., ground-state polymer conformations wrap the string completely. If the
attraction between a monomer and the string becomes stronger than the
interaction between stacked, neighboring monomer layers,
regular monolayer films surrounding the
string are formed, i.e., single-walled tubes with an ordered arrangement of
monomers. It is noticeable that there is a competition between
different chiral orientations of the wrapping in dependence of 
the monomer--string interaction length scale $\sigma_\mathrm{f}$~\cite{vogel10xxx}. This
behavior is in a similar manner known from carbon
nanotubes~\cite{dekker98nat}. Defining the wrapping vector $\vec{c}=n\vec{a}_1+m\vec{a}_2$,
with $\vec{a}_1$ and $\vec{a}_2$ being the unit vectors of the
structure, we even find 
the limiting
``armchair'' and ``zigzag'' structures, corresponding to $m=n$ and
$m=0$. Examples of nanotube-like polymer conformations with different chiralities are
shown for
$\sigma_\mathrm{f}=1.50,1.57$ in 
Figs.~\ref{fig:5}(a) and~\ref{fig:5}(b), 
respectively, and for $\sigma_\mathrm{f}=0.65$ in Fig.~\ref{fig:5}(c) (all at $\epsilon_\mathrm{f}=5$).
The alignment of monomers in Fig.~\ref{fig:5}(a) is almost parallel to the string, whereas the 
conformation in Fig.~\ref{fig:5}(b) exhibits a noticeable chiral winding. 
If the radius of the monolayer polymer tube does not allow for a perfect monomer alignment, defects occur
and cause the formation of structural domains with different chiralities within the
same conformation~\cite{vogel10xxx}, as in the example shown in Fig.~\ref{fig:5}(c).

To summarize, we have constructed the entire conformational phase
diagram of a hybrid system consisting of a flexible polymer
and an ultrathin attractive nanowire in dependence of the
energy scales and length scales associated to the polymer--nanowire interaction. 
We identified conformational phases of compact spherical polymer droplets 
inclosing or excluding the string, and a phase of compact but curved
shapes (crescent-shaped structures). For sufficiently large string attraction strengths, we 
observe the formation of cylindrical conformations
which in the extreme case of monolayer structures possess strong similarities 
to nanotubes. This is particularly interesting as it shows that polymers
can form tubelike structures in a controlled way. Since the polymer tube can adapt
any orientation of the guiding nanowire, also the formation of complex, nonlinear 
tube systems with bends is conceivable. This would enable a wide range 
of potential applications which are hard to construct by atomic nanotubes.

The authors would like to thank J.~Adler and T.~Mutat
from the Technion Haifa, for valuable discussions on
nanotubes. This project is supported by the J\"ulich/\break{}Aachen/Haifa
Umbrella program under Grant No.~SIM6. Supercomputer time is provided by
the Forschungszentrum J\"ulich under Project No.\ jiff39.
\end{document}